\documentclass[twocolumn]{aastex62}
\usepackage[mediumspace,mediumqspace,Grey,squaren]{SIunits}

\graphicspath{{./}{figures/}}

\submitjournal{ApJ}

\shorttitle{Extended main-sequence turnoffs of $h$ and $\chi$ Persei}
\shortauthors{C. Li et al.}
\begin{document}

\title{Extended main-sequence turnoffs in the double cluster $h$ and
  $\chi$ Persei: The complex role of stellar rotation}

\correspondingauthor{Chengyuan Li}
\email{chengyuan.li@mq.edu.au}

\author{Chengyuan Li} 
\affil{Department of Physics and Astronomy, Macquarie University,
  Sydney, NSW 2109, Australia}
\affil{Centre for Astronomy, Astrophysics and Astrophotonics,
  Macquarie University, Sydney, NSW 2109, Australia}
\affiliation{Key Laboratory for Optical Astronomy, National
  Astronomical Observatories, Chinese Academy of Sciences, 20A Datun
  Road, Beijing 100012, China}
  
\author{Weijia Sun}
\affiliation{Department of Astronomy, Peking University, Yi He Yuan Lu
  5, Hai Dian District, Beijing 100871, China}
\affiliation{Kavli Institute for Astronomy and Astrophysics, Peking
  University, Yi He Yuan Lu 5, Hai Dian District, Beijing 100871,
  China}

\author{Richard de Grijs}
\affiliation{Department of Physics and Astronomy, Macquarie
  University, Sydney, NSW 2109, Australia}
\affil{Centre for Astronomy, Astrophysics and Astrophotonics,
  Macquarie University, Sydney, NSW 2109, Australia}
\affiliation{International Space Science Institute--Beijing, 1
  Nanertiao, Zhongguancun, Hai Dian District, Beijing 100190, China}

\author{Licai Deng}

\affiliation{Key Laboratory for Optical Astronomy, National
  Astronomical Observatories, Chinese Academy of Sciences, 20A Datun
  Road, Beijing 100012, China}
\affil{Department of Astronomy, China West Normal University, Nanchong
  637002, China}
  
\author{Kun Wang} 
\affil{Department of Astronomy, China West Normal University, Nanchong
  637002, China} 
\affil{Instituto de Astrof\'{}sica de Canarias, 38200 La Laguna,
  Tenerife, Spain}

\author{Giacomo Cordoni} 
\affiliation{Dipartimento di Fisica e Astronomia `Galileo Galilei',
  Universit\`a di Padova, Vicolo dell'Osservatorio 3, IT-35122 Padova,
  Italy}

\author{Antonino P. Milone}
\affiliation{Dipartimento di Fisica e Astronomia `Galileo Galilei',
  Universit\`a di Padova, Vicolo dell'Osservatorio 3, IT-35122 Padova,
  Italy}

\begin{abstract}
Using {\sl Gaia} Data Release 2 photometry, we report the detection of
extended main-sequence turnoff (eMSTO) regions in the color--magnitude
diagrams (CMDs) of the $\sim 14$ Myr-old double clusters $h$ and
$\chi$ Persei (NGC 869 and NGC 884). We find that stars with masses
below $\sim$1.3 $M_{\odot}$ in both $h$ and $\chi$ Persei populate
narrow main sequences (MSs), while more massive stars define the
eMSTO, closely mimicking observations of young Galactic and Magellanic
Cloud clusters (with ages older than $\sim$30 Myr). Previous studies
based on clusters older than $\sim$30 Myr find that rapidly rotating
MS stars are redder than slow rotators of similar luminosity,
suggesting that stellar rotation may be the main driver of the
eMSTO. By combining photometry and projected rotational velocities
from the literature of stars in $h$ and $\chi$ Persei, we find no
obvious relation between the rotational velocities and colors of
non-emission-line eMSTO stars, in contrast with what is observed in
older clusters. Similarly to what is observed in Magellanic Cloud
clusters, most of the extremely rapidly rotating stars, identified by
their strong H$\alpha$ emission lines, are located in the red part of
the eMSTOs. This indicates that stellar rotation plays a role in the
color and magnitude distribution of MSTO stars. By comparing the
observations with simulated CMDs, we find that a simple population
composed of coeval stars that span a wide range of rotation rates is
unable to reproduce the color spread of the clusters' MSs. We suggest
that variable stars, binary interactions, and stellar rotation affect
the eMSTO morphology of these very young clusters.

\end{abstract}


\section{Introduction} \label{S1}

During the last decade, a large fraction of young and intermediate-age
massive clusters ($\leq$2 Gyr) in the Large and Small Magellanic
Clouds (LMC and SMC) have been found to exhibit extended main-sequence
turnoff (eMSTO) regions
\citep[e.g.,][]{Mack07a,Milo09a,Li14a,Li17a,Milo18a}. A similar
feature was recently detected in some Galactic open clusters (OCs) by
{\sl Gaia} \citep{Mari18a,Cord18a}.

A prevailing scenario to explain these eMSTOs suggests that turnoff
stars in these clusters are characterized by a wide range of rotation
rates \citep[e.g.,][]{Bast09a,Bran15a,Dant17a}. First, the effect of
gravity darkening caused by rapid stellar rotation would cause a
rapidly rotating star to have a lower surface temperature than its
non-rotating counterpart \citep{vonz24a}. In addition, individual
stars will also exhibit temperature gradients on their surfaces, so
that the observed colors and magnitudes depend on their inclination
with respect to the line of sight. Second, rotation may increase the
stellar main-sequence (MS) lifetime through expansion of the
convective shell, leading to transportation of shell material to the
stellar core \citep{Maed00a}. As a consequence, rapid rotation will
retain massive stars on the MS much longer than less massive stars,
thus populating the bright extension to the MS. { Finally, a range
  in stellar rotation rates will mimic an artificial age spread in a
  young cluster if we assume that all its member stars are
  non-rotating stars. The magnitude of this artificial age spread
  would be a function of the cluster's typical age, as has been
  confirmed for clusters in the Magellanic Clouds \citep{Nied15a} and
  for Galactic open clusters \citep{Cord18a}.}

Direct spectroscopic studies of individual stars in some young massive
clusters have shown that rapidly rotating stars are generally redder
than slowly or non-rotating stars \citep{Dupr17a,Mari18a,Mari18b}. In
addition, a survey of Be stars in young LMC clusters has confirmed
that Be stars, which are supposed to be rotating at near-critical
rotation rates, are systematically redder than normal MS stars
\citep{Milo18a}. Stars in Galactic OCs have been reported to exhibit a
large range of rotation rates \citep{Huang06a,Huang10a}. A number of
studies have explored how stars with different rotation rates will
populate their clusters' turnoff regions
\citep[e.g.,][]{Bran15b}. However, only a limited number of studies
have revealed direct correlations between the rotation rates exhibited
by turnoff stars and their distributions in cluster color--magnitude
diagrams (CMDs) \citep{Bast18a,Mari18a}.

In this paper, we study the turnoff and upper-MS stars of the double
clusters $h$ and $\chi$ Persei (NGC 869 and NGC 884). We aim to
explore the correlation, if any, between their rotational velocities
and the stellar color--magnitude distributions. We combine photometry
from {\sl Gaia} Data Release 2 \citep[DR2;][]{Gaia16a,Gaia18a} and the
stellar catalog of \cite{Huang06a,Huang10a}, which includes
information about the rotation rates of individual stars in our sample
clusters.

\section{Data Reduction} \label{S2}

We obtained the photometric stellar catalog from the {\sl Gaia}
archive\footnote{\url{http://gea.esac.esa.int/archive/}}. Based on the
parameters provided by \cite{Khar13a}, we adopted radii of,
respectively, 3,050 arcsec and 3,500 arcsec as the areas to search for
stars associated with $h$ and $\chi$ Persei. These areas are slightly
larger than their tidal radii. The resulting stellar catalog contains
143,900 and 187,596 stars associated with the regions occupied by $h$
and $\chi$ Persei, respectively, for which we have information about
(1) the spatial coordinates (right ascension and declination), (2)
parallaxes, (3) proper motions, and (4) photometry (in the $G_{\rm
  bp}$, $G_{\rm rp}$, and $G$ passbands). We first removed all stars
associated with unrealistic negative parallaxes. These stars may have
been affected by confusion in the observation-to-source matching, and
they will have unreliable proper motions \citep{Gaia18a}. This step
led to the removal of $\sim$32\% of the stars from the raw stellar
catalog. Because $h$ and $\chi$ Persei are located in close vicinity
to each other, their cluster regions as defined by their tidal radii
overlap. We further removed 77,227 stars that appeared as duplicates
in both clusters' stellar catalogs and combined them into a common
catalog. Finally, 147,748 stars were located in the area containing
both clusters.

\begin{figure*}
\includegraphics[width=2\columnwidth]{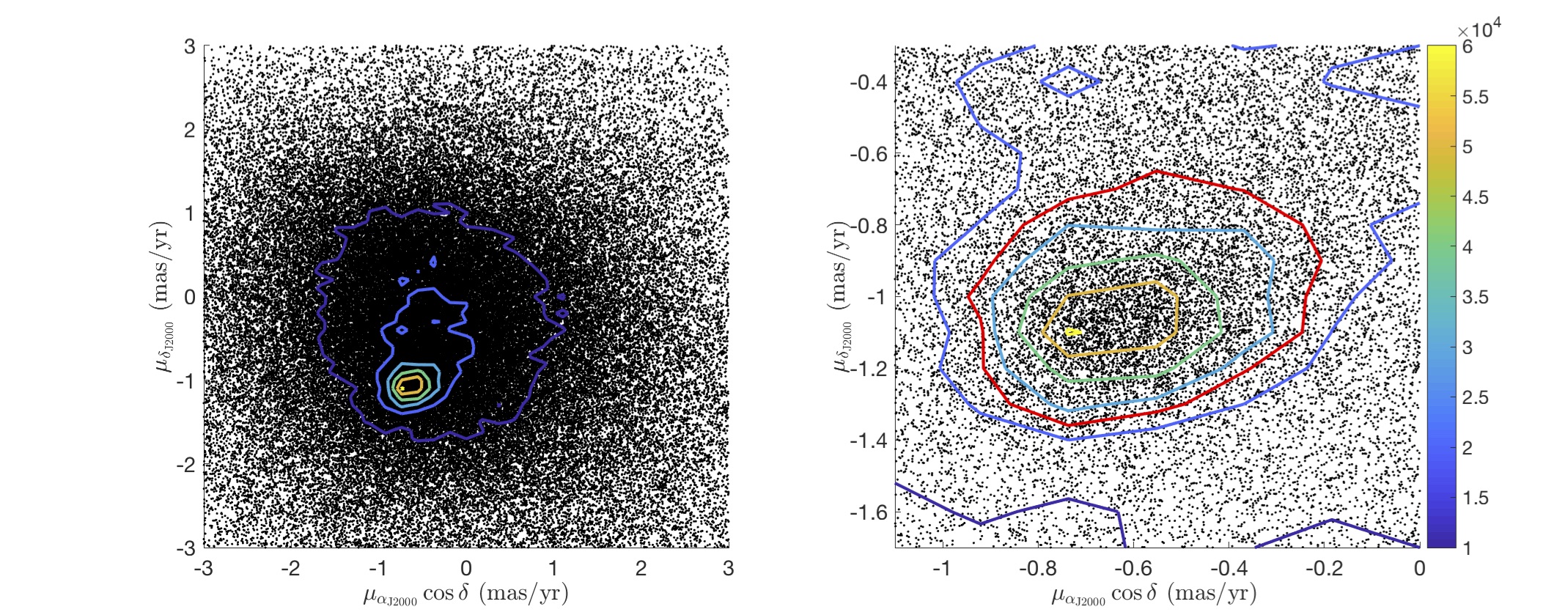}
\caption{Proper-motion diagrams of stars in the search area for the
  clusters $h$ and $\chi$ Persei. Overplotted are contours
  representing stellar number densities (in units of $N$ deg$^{-2}$).
  The right-hand panel exhibits a zoomed-in region. 
  Stars located in the red loop
  (i.e., the isodensity loop for 25,000 yr$^2$ mas$^{-2}$) were
  selected as candidate cluster stars.}
\label{F1}
\end{figure*}

To decontaminate the field stars, we explored the proper motion map of
all stars. We calculated their density contours across the entire
proper-motion diagram. We found that there is a clearly overdense
region centered at about $\mu_{\alpha_{\rm
    J2000}}\cos{\delta}=$$-$0.65 mas yr$^{-1}$ and $\mu_{\delta_{\rm
    J2000}}=$$-$1.05 mas yr$^{-1}$, which corresponds to the bulk
motions of both $h$ and $\chi$ Persei, where $\mu_{\alpha_{\rm
    J2000}}$ and $\mu_{\delta_{\rm J2000}}$ are the angular velocities
of the stellar proper motions in the right ascension and declination
directions, and $\delta$ is the declination. We did not find any
significant differences between the proper motions of $h$ and $\chi$
Persei. Our results are well-matched with literature results for $h$
and $\chi$ Persei \citep{Cant18a}. The proper-motion map for all
stars, as well as the calculated number density contours, are shown in
Figure \ref{F1}. We selected all stars located within the area
enclosed by the isodensity contour representing 25,000 yr$^2$
mas$^{-2}$ (see the red line in Figure \ref{F1}) as candidate cluster
stars. This selection removed 95\% of all detected stars, leaving only
7,452 stars.

\begin{figure}
\includegraphics[width=\columnwidth]{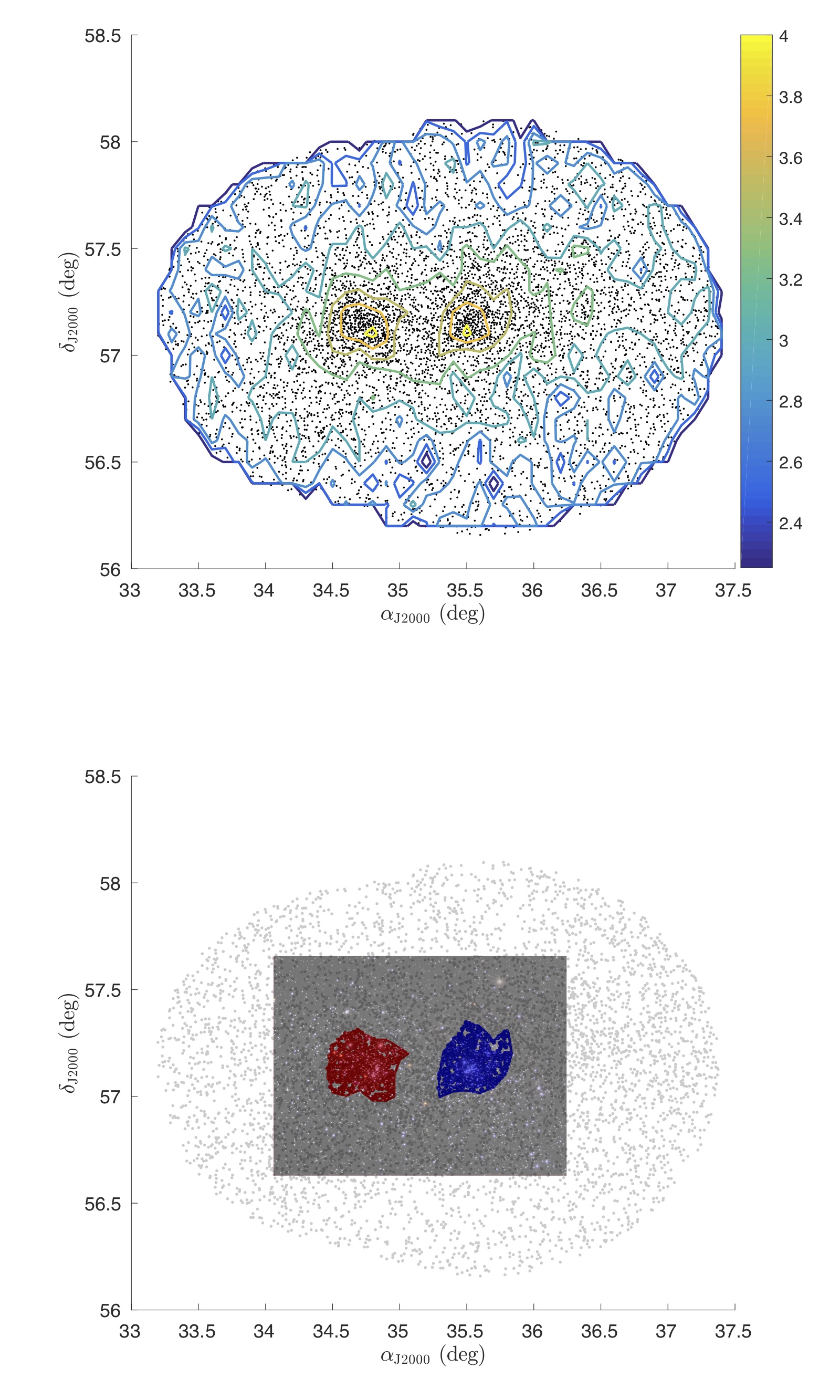}
\caption{(top) Spatial distributions of the proper-motion-selected
  stars in our search area, as well as their number density contours
  ($\log{(N {\rm deg^{-2}})}$). (bottom) Spatial distribution of all
  selected stars, superikposed on the real color image background
  (N. A. Sharp/NOAO/AURA/NSF). Stars associated with $h$ and $\chi$
  Persei are highlighted by red and blue dots, respectively.}
\label{F2}
\end{figure}

In Figure \ref{F2} we present the spatial distribution of all 7,452
stars in our final selection. They exhibit two clear clumps centered
at $\alpha_{\rm J2000}\sim34.75$ deg (2$^{\rm hh}$19$^{\rm
  mm}$0.12$^{\rm ss}$), $\delta_{\rm J2000}\sim57.15$ deg
(57$^{\circ}$9$'$0.00$''$) and $\alpha_{\rm J2000}\sim35.52$ deg
(2$^{\rm hh}$22$^{\rm mm}$4.80$^{\rm ss}$), $\delta_{\rm
  J2000}\sim57.14$ deg (57$^{\circ}$8$'$24.00$''$). These results are
generally consistent with those of \cite{Wu09a}. We plot their number
density contours by assigning all stars to 1,196 spatial grid cells
(46$\times$26 cells in [$\alpha_{\rm J2000}$, $\delta_{\rm J2000}$]
space) and calculated the average stellar number density in each
cell. At the edge of the region, the average stellar number density is
$\sim$56,000 deg$^{-2}$, while in the central region, the isodensity
contour associated with a stellar number density of
$\sim$3$\times$10$^5$ deg$^{-2}$ shows two distinct closed loops,
indicating that both clusters are clearly separated at this number
density level. We thus selected stars located within these two loops
(see the red and blue lines in Figure \ref{F2}) to represent the $h$
and $\chi$ Persei subsamples. Thus, we identified 828 and 720 stars in
$h$ and $\chi$ Persei, respectively, as shown in Figure \ref{F2} (red
and blue dots, respectively).

OCs located in the dusty Galactic disk may be severely affected by
reddening. Spatially variable extinction (differential reddening) will
introduce uncertainties in the actual colors and magnitudes of member
stars located in different parts of an OC, thus potentially mimicking
an artificial eMSTO. However, only a small number of stars brighter
than $G= 17$ mag have associated reddening values in the {\sl Gaia}
DR2 photometric catalog. Therefore, we used the Galactic Dust
Reddening and Extinction service provided by the NASA/IPAC Infrared
Science
Archive\footnote{\url{https://irsa.ipac.caltech.edu/applications/DUST/}}
\citep{Schl11a} to correct our cluster photometry for differential
reddening effects. We aim to correct all 7,452 stars for differential
reddening. Because the typical spatial resolution of the Galactic Dust
Reddening and Extinction service is $5'$, we examined the average
extinction, $E(B-V)$, for 16$\times$17 adjacent circular regions with
radii of $5'$ in [$\alpha_{\rm J2000}$, $\delta_{\rm J2000}$] space.

\begin{figure}
\includegraphics[width=\columnwidth]{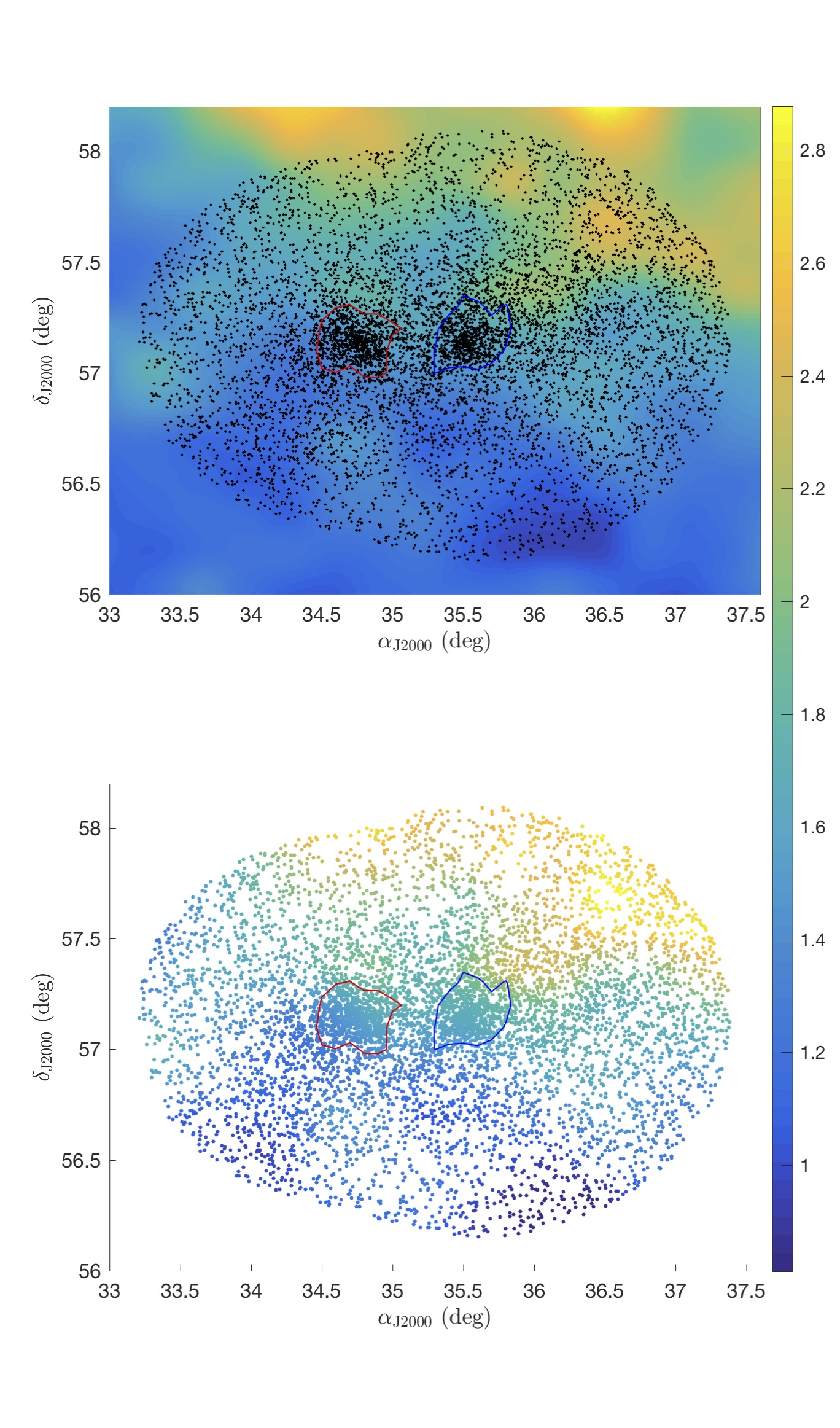}
\caption{(top) Spatial distributions of stars overplotted on the
  reddening atlas, as indicated by the color bar (in mag). (bottom)
  Spatial distributions of stars with their interpolated reddening
  values color-coded. Red and blue loops are the boundaries defined
  for the subsamples coinciding with $h$ and $\chi$ Persei,
  respectively.}
\label{F3}
\end{figure}

The reddening is higher in the north and east and lower in the south
and west, as shown in Figure \ref{F3} (indicated by the color-coded
background). We confirmed this apparent reddening gradient by visually
inspecting an Infrared Astronomical Satellite (IRAS)
\unit{100}{\micro\meter}
image\footnote{\url{https://irsa.ipac.caltech.edu/Missions/iras.html}}
as well. We then calculated the individual stellar extinction values
by matching their spatial coordinates with the derived reddening map.
All cluster stars are subject to reddening ranging from $A_{V} =0.98$
mag to $A_{V} =2.41$ mag; 95\% (7,164 stars) have reddening values
between $A_{V}=1.10$ mag and $A_{V}=2.27$ mag, with an average
reddening value of $A_{V}=1.59$ mag. As regards the $h$ and $\chi$
Persei subsamples, their average reddening and dispersion (covering
95\% of all stars) are $A_{V} =1.49_{-0.14}^{+0.15}$ mag and
$A_{V}=1.60_{-0.11}^{+0.18}$ mag, respectively. $\chi$ Persei is
subject to higher average reddening than $h$ Persei, which is expected
given its location on the eastern side of $h$ Persei since the IRAS
\unit{100}{\micro\meter} image has revealed that this region is very
dusty.

\begin{figure*}
\includegraphics[width=2\columnwidth]{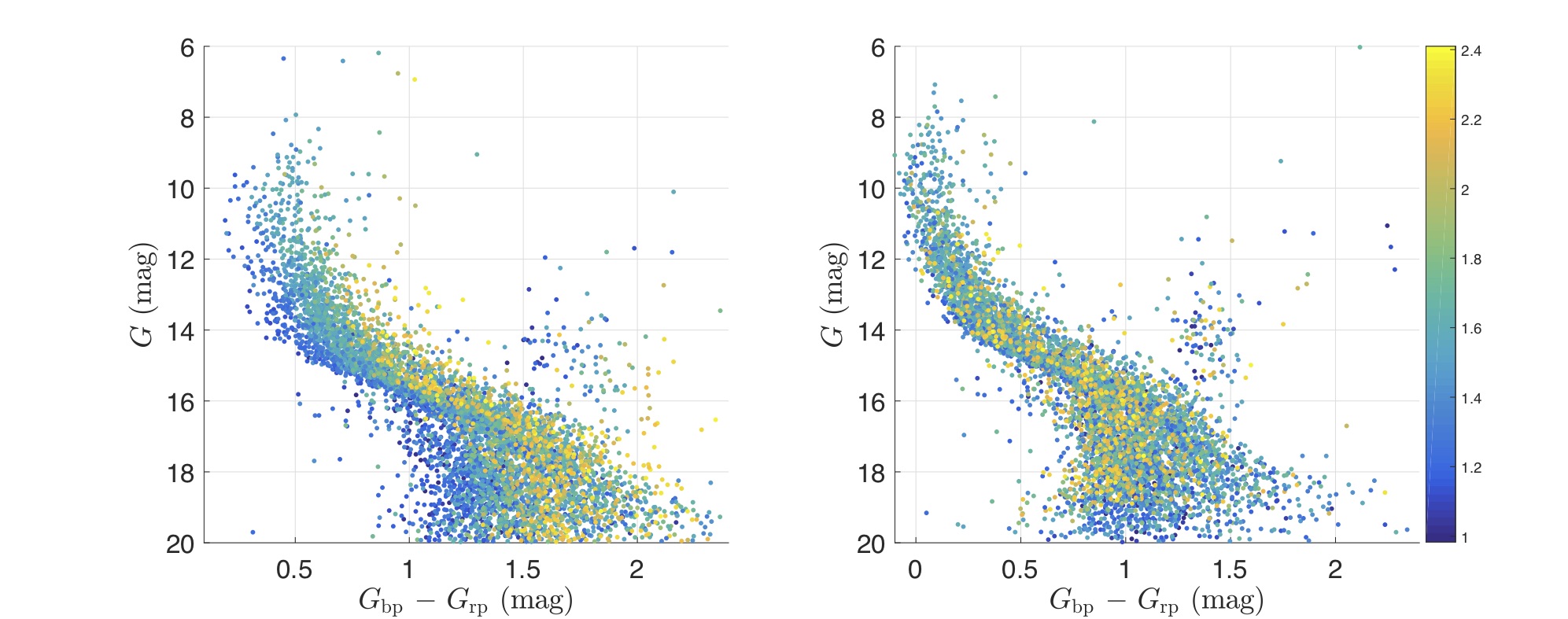}
\caption{(left) Raw CMD of all proper-motion-selected stars in the
  search area. (right) Differential reddening corrected CMD for all
  stars. The color bar indicates the interpolated reddening for
  individual stars.}
\label{F4}
\end{figure*}

Finally, we corrected our stellar photometry to represent the average
reddening using the \cite{Card89a} and \cite{Odon94a} extinction curve
with $R_{V}=3.1$. In Figure \ref{F4} we show the CMDs of all stars
before (left) and after (right) correcting for differential
reddening. A clear color gradient is seen across the entire CMD if
differential reddening is not corrected for, with stars affected by
higher reddening being systematically redder than stars affected by
less reddening. After correction, this color gradient all but
disappears.

\section{Results}

\begin{figure*}
\includegraphics[width=2\columnwidth]{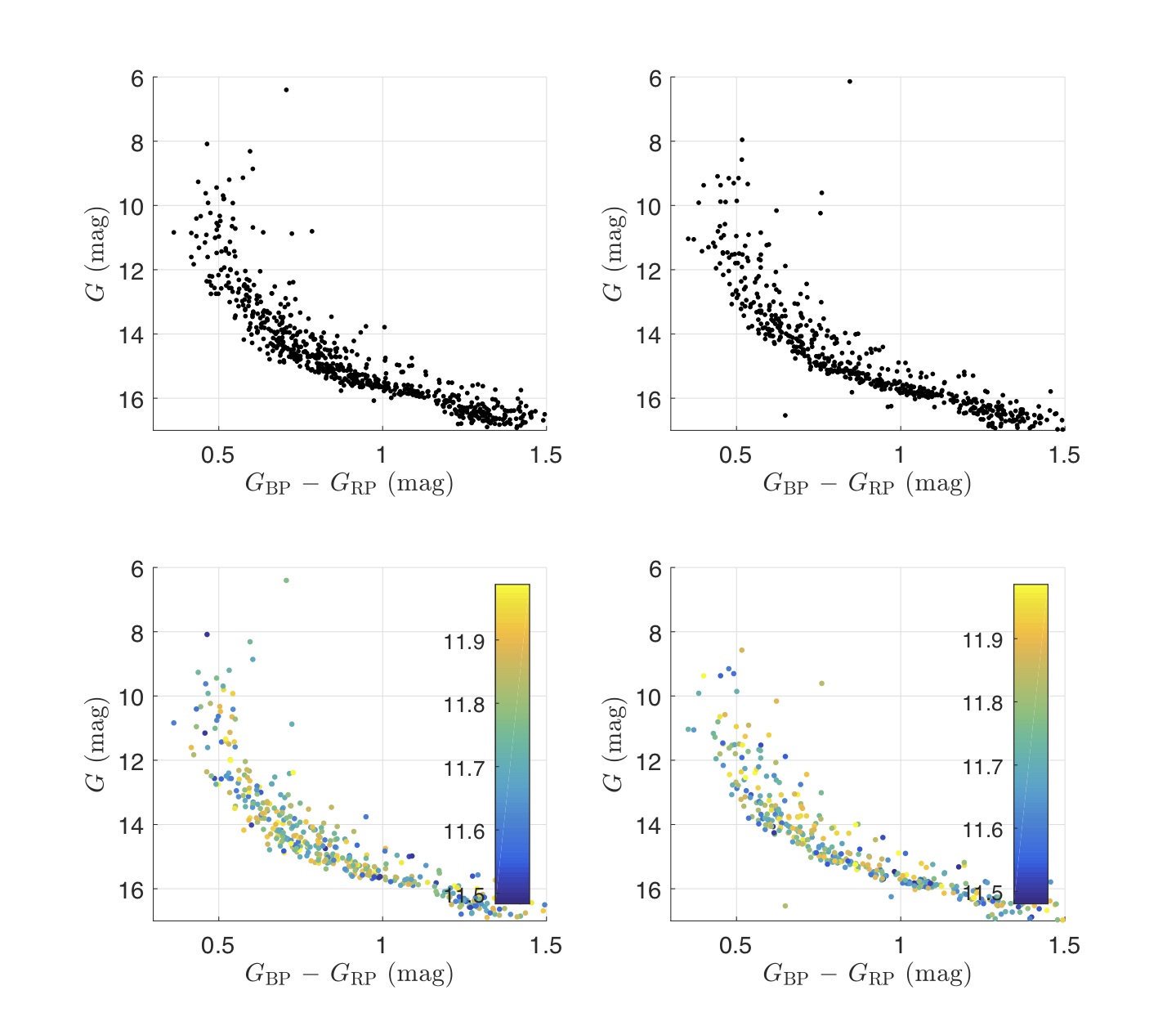}
\caption{Differential-reddening-corrected CMDs composed of stars in
  $h$ and $\chi$ Persei (left and right, respectively). The
  corresponding bottom panels show the distance-constrained CMDs, as
  indicated by the color bars.}
\label{F5}
\end{figure*}

The differential-reddening-corrected CMDs of $h$ and $\chi$ Persei are
presented in Figure \ref{F5}. Our results exhibit clear similarities
to the CMDs of young clusters in the Magellanic Clouds
\citep[e.g.,][]{Milo15a,Milo16a,Milo18a}. Stars fainter than $G\sim
15.5$ mag populate a narrow MS and a clear binary envelope, while the
upper MS and MSTO exhibit clear color spreads which make the binary
envelope less distinct. There is an internal color--magnitude spread
for stars with $G\leq 15.5$ mag. These stars map onto a complex
morphology of the upper MS and the MSTO regions in the CMDs of $h$ and
$\chi$ Persei.

Because differential stellar distances may also broaden the MS, we
calculated the individual stellar distance moduli based on parallax
measurements. The average distance to $h$ and $\chi$ Persei was
previously reported as about 2.3 kpc \citep{Khar13a}. In the bottom
panel of Figure \ref{F5}, we plot the corresponding CMDs for all stars
with distances between 2.0 and 2.6 kpc. This distance range
corresponds to a parallax spread of $\pm$0.05 mas. However, we
emphasize that this selection might be too strict for the stars of
interest, since the typical parallax uncertainty for stars of $G\sim
15.5$ mag is 0.04--0.1 mas \citep{Gaia18a}. As shown in Figure
\ref{F5}, even though we have set strict constraints on the distance
range of our stellar sample, the extended MSs and MSTOs are still
obvious. No obvious gradient in mean magnitude for stars at different
distances was detected in either of these CMDs.

We next used the MESA Isochrone and Stellar Tracks
\citep[MIST][]{Paxt11a,Paxt13a,Paxt15a,Choi16a,Dott16a} models to
generate different isochrones to fit the observations. \citep{Sles02a}
derived an age of $\sim$13 Myr and solar metallicity for both $h$ and
$\chi$ Persei. Our fits agree with their results. To describe the blue
boundary and the knee of the MS ($G \sim 15$--16 mag), the
best-fitting isochronal age and metallicity for both $h$ and $\chi$
Persei are 13.8 Myr and solar metallicity ($Z=0.0142$ in the MIST
models), respectively. Using the parallaxes from {\sl Gaia} DR2, we
recalculated the average distance moduli for both $h$ and $\chi$,
resulting in $(m-M)_0 =11.69$ mag and 11.81 mag, respectively. The
adopted average reddening values for the clusters are $A_{V} =1.49$
mag and $A_{V}=1.60$ mag, respectively. We assumed that the average
stellar rotation rate was close to zero.

\begin{figure*}
\includegraphics[width=2\columnwidth]{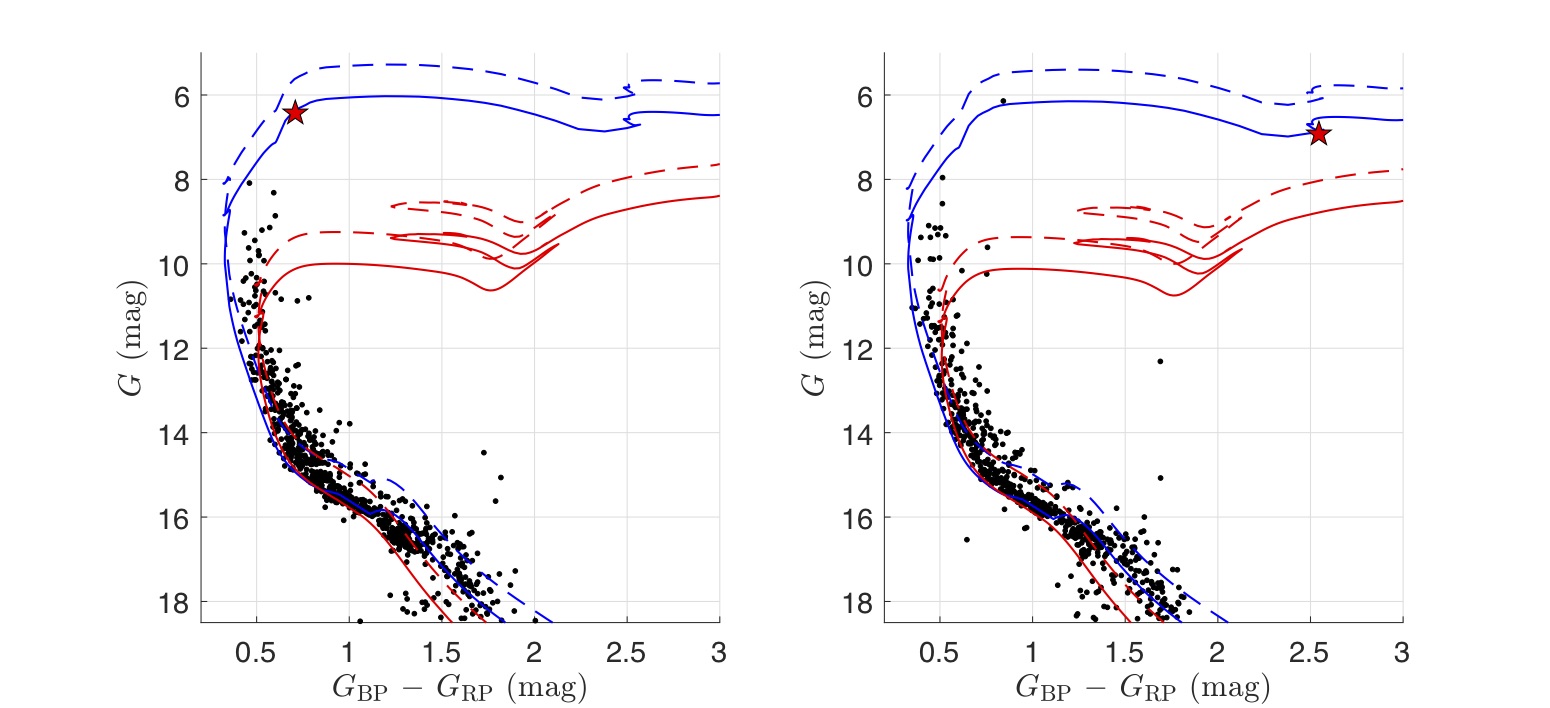}
\caption{Distance-constrained, differential-reddening-corrected CMDs
  of stars in $h$ and $\chi$ Persei (left and right, respectively),
  fitted by isochrones with ages of 14 Myr (blue solid lines) and 100
  Myr (red solid lines). The dashed lines are the corresponding loci
  for unresolved, equal-mass binaries. Red stars are the A-type
  turnoff star ($h$ Persei) and the M supergiant star ($\chi$ Persei)
  used to determine the best fits of the younger isochrones.}
\label{F6}
\end{figure*}

This age of the best-fitting isochrone can be constrained by an A-type
star located in the red portion of the Hertzsprung gap of $h$ Persei
and an M supergiant star in $\chi$ Persei (see the red pentagrams in
Figure \ref{F6}). However, these isochrones only describe the blue
edge of the upper MSs and the MSTOs of $h$ and $\chi$ Persei. They do
not describe the red part of the MSs, as expected. The color spreads
of stars in the upper MS and MSTO regions of $h$ and $\chi$ Persei
(for $G\leq 15.5$ mag) are about $\Delta(B_{\rm bp}-B_{\rm rp})\sim
0.25$ mag. The average photometric uncertainties for stars in this
magnitude range are $\delta{B_{\rm bp}}\leq0.006$ mag and
$\delta{B_{\rm rp}}\leq0.004$ mag \citep[their Figures 10 and
  11]{Evan18a}, which translates into a color uncertainty of
$\delta{(B_{\rm bp}-B_{\rm rp})}\leq0.007$ mag. The observed color
spread is about 35 times the theoretical color uncertainty. The
possibility that the extended upper MS and MSTO regions are caused by
photometric uncertainties is therefore negligible. The extended upper
MS is unlikely to have been caused by unresolved binaries either,
since the upper range of the MS is almost vertical in the CMD. In
Figure \ref{F6}, we present the observed CMDs of $h$ and $\chi$ Persei
and the best-fitting isochrones, as well as the loci of the equal-mass
unresolved binaries.

To quantify the width of the clusters' eMSTO regions, we next obtained
fits using older isochrones characterized by the same metallicity and
extinction to describe the red part of the MSs. We found that even an
older isochrone with an age of $\sim$100 Myr (see the red lines in
Figure \ref{F6}) cannot satisfactorily describe the morphology of the
upper MS. These isochrones (with ages of $\sim$14 Myr and 100 Myr) can
only reproduce the MS for $G\leq 12$ mag, while the observed MSs show
clear color spreads at $G\leq 15$ mag. If these color spreads are
caused by genuine age spreads, we should expect well-populated red
giant branches in both young clusters, but this is not observed. In
summary, although fits of single stellar populations to the
observations are inadequate, the clusters' eMSTO regions and broadened
MSs are unlikely caused by significant age spreads among their member
stars.

An alternative explanation, prevailing for most young clusters
exhibiting eMSTO regions \citep[e.g.,][however, see
  \cite{Goud17a,Goud18a,Piat17a}]{Nied15a,Dant17a,Dupr17a,Mari18a},
invokes stellar rotation. This scenario suggests that the
color--magnitude spread of MS and MSTO stars may be caused by
differences in the cluster stars' rotation rates. { Specifically,
  the gravity darkening caused by different stellar rotation rates and
  inclinations will render different stellar colors and
  luminosities. Rotational mixing would prolonge the MS lifetimes of
  rapidly rotating stars, resulting in MSTO stars of different masses
  \citep{Yang11a}. All of these effects cause stellar rotation to play
  a complex role in shaping the morphology of a cluster's MSTO
  region.}

To address whether stellar rotation is the underlying cause of the
eMSTOs of $h$ and $\chi$ Persei, a direct comparison with models is
required. To do so, we requested synthetic photometry representative
of a simple stellar population from the SYCLIST model
suite\footnote{\url{https://www.astro.unige.ch/syclist/index/Formats/\#cluster}}
\citep{Geor14a} for different rotation rates. The requested model
contains 5000 single stars and 5000 unresolved binary systems at a
fixed age of 14 Myr ($\log{t/{\rm yr}} =7.15$) and for solar
metallicity ($Z=0.014$), characterized by the empirical stellar
rotational velocity distribution of \cite{Huang10a}. For rapidly
rotating stars, both gravity darkening and limb darkening are
considered \citep{Espi11a}. Photometric data points for our simulated
cluster are given in the standard Johnson--Cousins $UBVRI$ photometric
system. Using color--color mapping between the Johnson--Cousins
passbands and the {\sl Gaia} photometric system provided by the {\sl
  Gaia} DR2
documentation\footnote{\url{http://gea.esac.esa.int/archive/documentation/GDR2/}},
we calculated the corresponding {\sl Gaia} photometry in the $G$,
$G_{\rm bp}$, and $G_{\rm rp}$ passbands.

\begin{figure*}
\includegraphics[width=2\columnwidth]{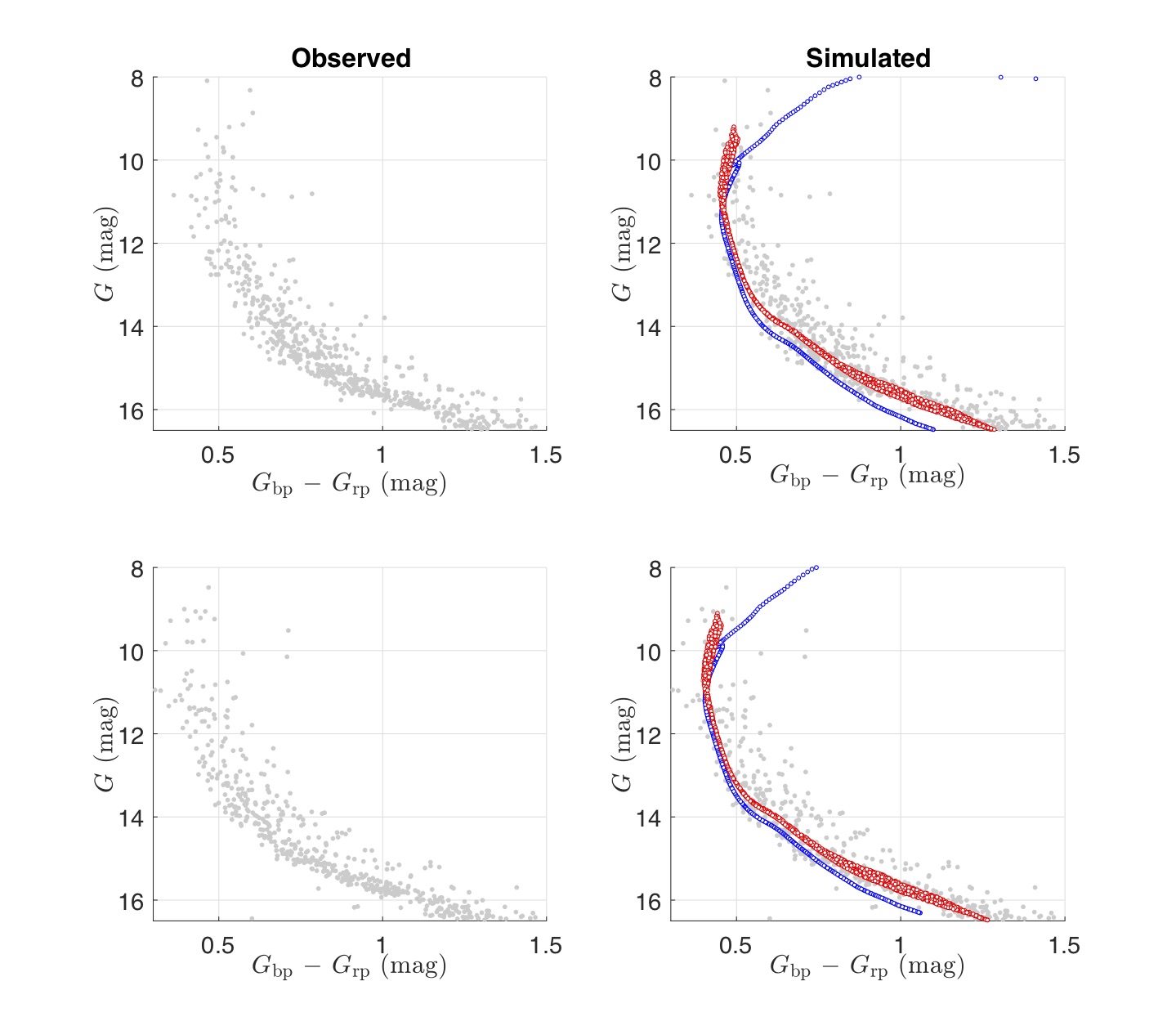}
\caption{Observed CMDs of (top left) NGC 869 and (bottom left) NGC
  884, as well as (right) the corresponding simulated CMDs pertaining
  to non-rotating (blue) and rapidly rotating (for
  $\omega=0.9\omega_{\rm cr}$, red) populations.}
\label{F7a}
\end{figure*}

Similarly as in \cite{Cord18a}, in Figure \ref{F7a} we show the CMDs
of the observed cluster stars as well as a set of simulated CMDs for
non-rotating and rapidly rotating populations. For clarity, the
simulated stars do not contain any unresolved binaries and their
photometry is noise-free. Although the rapidly rotating stars exhibit
some spread in their color--magnitude distributions, the observed
widths of the MSTO regions are still much wider than those resulting
from the simulations.

For a more realistic assessment, we modeled a synthetic stellar
population characterized by various stellar rotation rates and added
realistic observational noise to each star's photometry to mimic the
effects of both photometric uncertainties and possible differential
reddening residuals. We added similar photometric uncertainties to our
stellar photometry as in the observations. Although we corrected the
individual stellar photometry for differential reddening, we are aware
of the fact that a residual reddening spread may introduce a
color--magnitude spread in the CMD. Indeed, even though we added
realistic photometric uncertainties to our simulation, the simulated
MS is still narrower than the observed MSs for $G\geq15.5$ mag. These
additional MS spreads may be caused by reddening residuals. Finally,
we included additional photometric uncertainties of $\delta{G}\sim
0.006$ mag and $\delta{(G_{\rm bp}-G_{\rm rp})}\sim 0.009$ mag to our
simulation to reproduce the spread caused by this residual
differential reddening.

\begin{figure*}
\includegraphics[width=2\columnwidth]{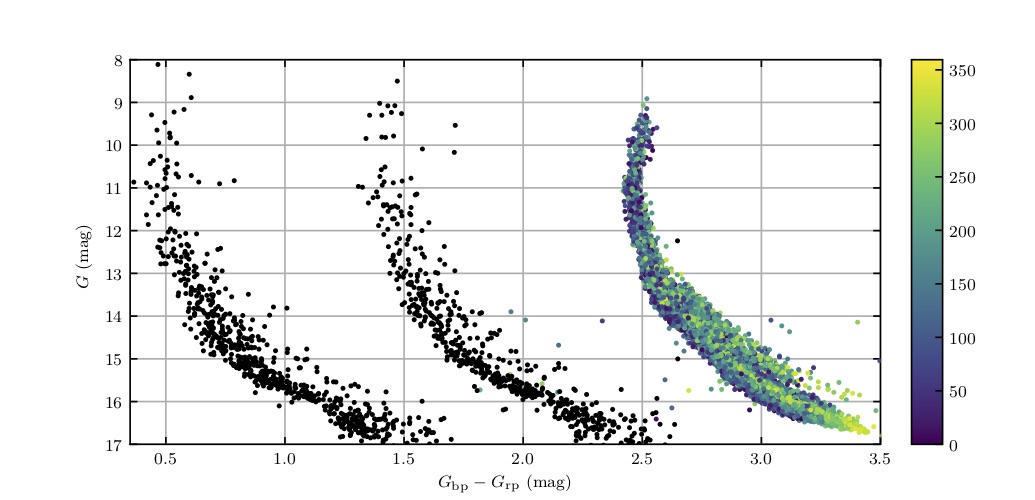}
\caption{CMDs of our observations (black circles) and the synthetic
  cluster (colored dots), with their $V\sin{i}$ indicated by the color
  bar (in km s$^{-1}$). To visually highlight the similarities and
  differences between the observations and the synthetic cluster, NGC
  884 (middle) and the model cluster (right) have been shifted in
  color.}
\label{F7b}
\end{figure*}

Figure \ref{F7b} shows the CMDs of both the synthetic clusters and the
observations. For the synthetic clusters, we use a range in color to
represent their projected rotational velocities ($V\sin{i}$). For a
stellar population of only 14 Myr old, the color distribution of the
MSTO stars does not depend on their rotational velocities. This is
different from the situation for the clusters M11 and NGC 2818, as
reported by \cite{Mari18a} and \cite{Bast18a}, whose upper MS and MSTO
stars show a clear correlation between their colors and the projected
rotational velocities: the rapid rotators preferentially occupy the
red parts of the MSTO regions. This is so, because rotational mixing
has prolonged the MS lifetimes of the rapidly rotating stars, which
masks the effects of gravity darkening. In turn, this causes their
color--magnitude distribution to define an MSTO locus that is almost
indistinguishable from that of the non-rotating stars. In addition,
the model implies that the color spread of stars with different
rotational velocities is still too small to reproduce the
observations. This is because for a stellar population of only 14 Myr
old, the upper MS and MSTO stars are too hot to show clear color
differences owing to gravity darkening when comparing extremely
rapidly rotating stars with their non-rotating counterparts.

We used the stellar catalogs provided by \cite{Huang06a} and
\cite{Huang10a}. These catalogs contain 89 and 95 stars located in the
areas of, respectively, $h$ and $\chi$ Persei, with their projected
rotational velocities calculated by comparison of moderate-resolution
observational spectra with synthetic spectra calculated by TLUSTY and
SYNSPEC \citep{Hube95a}. Specifically, they fitted the observed
absorption line profiles to the synthetic rotational profiles across a
grid of projected rotational velocities \citep[see Figure 1 of][as an
  example]{Huang06a}. We cross-matched these stars with our
proper-motion-selected samples. However, we remind reader that some of
their sample objects are located beyond the cluster regions we have
adopted. Nevertheless, we still use these stars in order to attain a
larger sample. We selected, respectively, 56 and 64 stars in common
for $h$ and $\chi$ Persei from both our catalog and those of
\cite{Huang06a} and \cite{Huang10a}. The latter stellar catalogs also
include information about double-lined spectroscopic binaries, radial
velocity variables (possible binaries), and H$\alpha$ emission-line
(Be) stars. Stars with unusual line profiles were removed because the
measurements of their projected rotational velocities may be
unreliable. Finally, 40, 11, and three stars were identified as normal
stars, binaries (including both confirmed double-lined spectroscopic
binaries and radial velocity variables), and emission-line stars in
$h$ Persei, while for $\chi$ Persei the corresponding numbers are 39,
16, and five.

\begin{figure*}
\includegraphics[width=2\columnwidth]{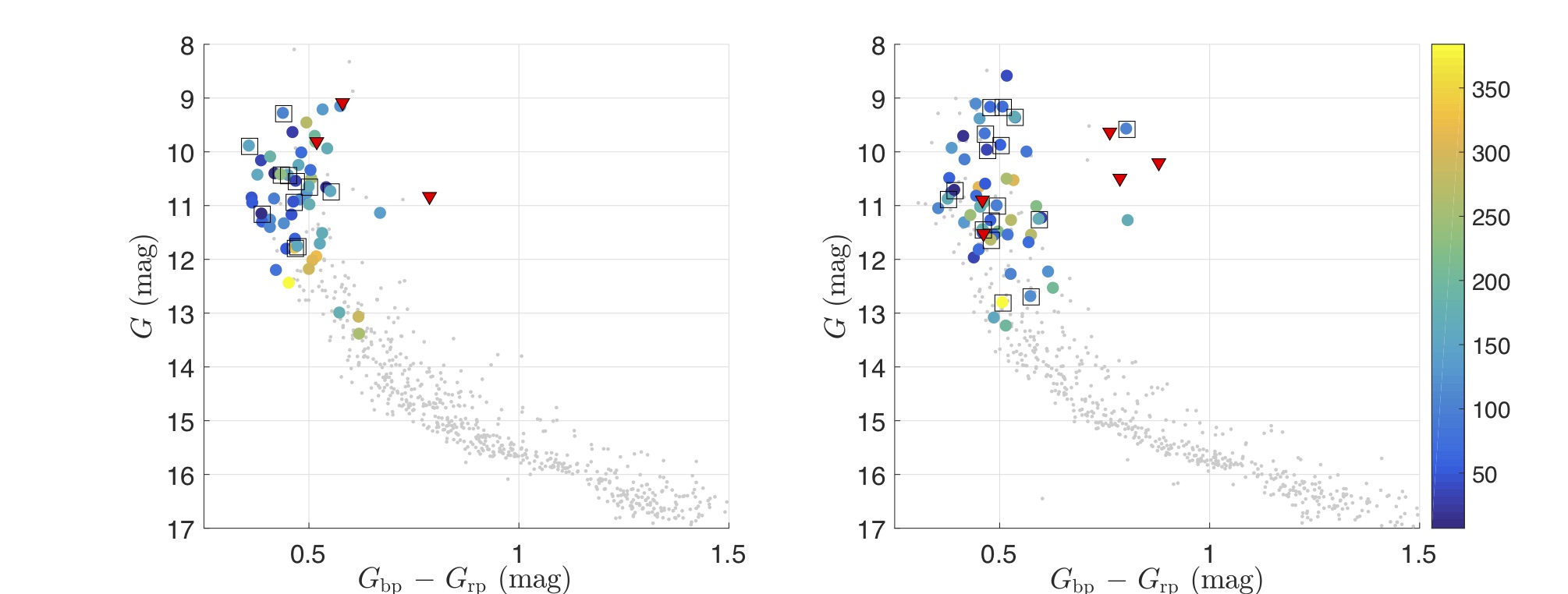}
\caption{Color--magnitude distributions of stars with measured
  projected rotational velocities, $V\sin{i}$, as indicated by the
  color bar (in km s$^{-1}$). Double-lined spectroscopic binaries and
  radial velocity variables (possible binaries) are highlighted in
  squares. Red triangles are stars showing emission lines.}
\label{F8}
\end{figure*}

In Figure \ref{F8} we show the CMDs of $h$ and $\chi$ Persei, along
with the measured $V\sin{i}$ values for the subsamples' normal stars
and binaries (enclosed in squares), as indicated by the color
bar. Indeed, no obvious correlation between the projected rotational
velocities and the color--magnitude distributions was detected in
either cluster. However, most of the Be stars are located in the red
part of the MS (see the red triangles in Figure \ref{F8}). This is
similar to the case for young clusters in the LMC
\citep{Milo18a}. Since Be stars are thought to be extremely rapidly
rotating stars with near-critical rotation rates, the fact that Be
stars are systematically redder than normal stars may indicate that
stellar rotation still plays a role in the eMSTO regions of $h$ and
$\chi$ Persei.

\begin{figure}
\includegraphics[width=\columnwidth]{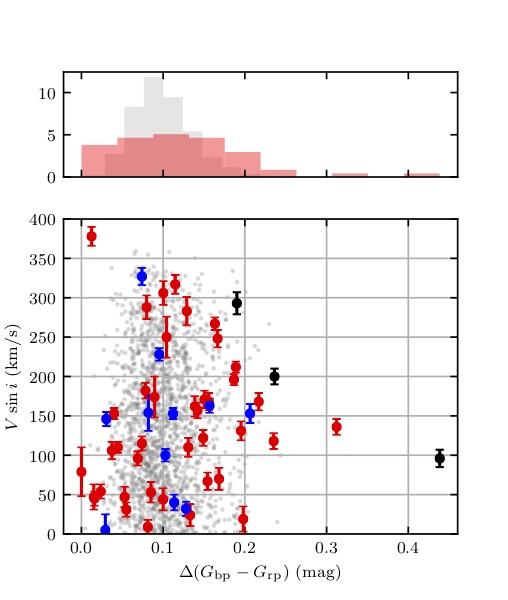}
\caption{(top) Observed $\Delta(G_{\rm bp}-G_{\rm rp})$ distribution
  (red) and corresponding distribution for the simulated CMD (grey).
  (bottom) $\Delta{(G_{\rm bp}-G_{\rm rp})}$--$V\sin{i}$ diagram for
  all stars in NGC 869 with projected rotational velocity
  measurements, where the red, blue, and black samples represent
  normal stars, binaries, and possible binaries, as well as
  emission-line stars. Circles and squares represent stars associated
  with $h$ and $\chi$ Persei, respectively. The background grey dots
  represent the same distribution for the synthetic cluster. The
  calculated Spearman's correlation coefficient between
  $\Delta{(G_{\rm bp}-G_{\rm rp})}$ and $V\sin{i}$ is 0.22.}
\label{F9a}
\end{figure}

\begin{figure}
\includegraphics[width=\columnwidth]{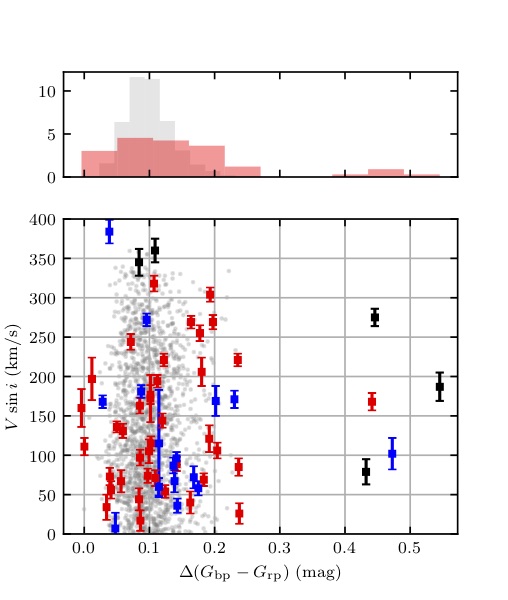}
\caption{As \ref{F9a}, but for NGC 884. The corresponding Spearman's
  coefficient is only 0.04}
\label{F9b}
\end{figure}

Figures \ref{F9a} and \ref{F9b} show the $\Delta{(G_{\rm bp}-G_{\rm
    rp})}$--$V\sin{i}$ distributions for all of our stars. Here,
$\Delta{(G_{\rm bp}-G_{\rm rp})}$ is the stellar color deviation with
respect to the zero-age MS (ZAMS) described by the best-fitting, 14
Myr-old isochrones. Stars located in the middle of the upper MS
exhibit the largest velocity dispersion. A major fraction of stars
located in the red portion of the ZAMS, with $\Delta{(G_{\rm
    bp}-G_{\rm rp})} \sim 0.07$--0.15 mag, cover the full spectrum of
projected rotational velocities, ranging from zero to $\sim$360 km
s$^{-1}$. Most stars located close to the ZAMS, with $\Delta{(G_{\rm
    bp}-G_{\rm rp})} \leq 0.07$ mag, have projected rotational
velocities of less than 200 km s$^{-1}$. Only two exceptions exhibit
projected rotational velocities of $\sim$400 km s$^{-1}$. Stars
located in the red part of the MS, with $\Delta{(G_{\rm bp}-G_{\rm
    rp})} \sim 0.15$--0.25 mag, exhibit a distribution of projected
rotational velocities from almost zero to $\sim$300 km s$^{-1}$. Seven
stars have $\Delta{(G_{\rm bp}-G_{\rm rp})} \geq 0.25$ mag, four of
which are identified as emission-line stars. There are no significant
differences between the distributions of possible binaries (blue
circles and squares) and normal stars (red circles and
squares). Similarly to \cite{Mari18a}, we calculated the Spearman's
correlation coefficient between $\Delta{(G_{\rm bp}-G_{\rm rp})}$ and
$V\sin{i}$. \cite{Mari18a} found a clear correlation between the
colors and the projected rotational velocities of eMSTO stars in M11,
characterized by a Spearman's correlation coefficient of $c=0.55$. For
NGC 869 and NGC 884 we only find $c=0.22$ and 0.04, respectively.
 
Using the same method as applied to the observations, for each
simulated star we calculated $\Delta{(G_{\rm BP}-G_{\rm rp})}$ with
respect to the 14 Myr-old isochrone. For the observations, the
best-fitting isochrone describes the blue edge of the eMSTO (Figure
\ref{F6}), while for the simulation, the isochrone is of course the
ridge-line. This means that the zeropoint for the observed
$\Delta{(G_{\rm bp}-G_{\rm rp})}$ distribution is different from that
for the simulation. After having corrected the zeropoint difference to
coincide with the observations, we compared the $\Delta{(G_{\rm
    BP}-G_{\rm rp})}$--$V\sin{i}$ distributions. We only examined the
1,785 simulated stars with $G\leq 13.5$ mag. None have $\Delta{(G_{\rm
    BP}-G_{\rm rp})}\ge 0.25$ mag, while more than 91\% have
$\Delta{(G_{\rm BP}-G_{\rm rp})}\leq 0.15$ mag. For comparison,
4--10\% of stars have $\Delta{(G_{\rm BP}-G_{\rm rp})}\ge 0.25$ mag (2
of 56 and 7 of 64 in NGC 869 and NGC 884, respectively). In addition,
34--36\% (19 of 56 and 23 of 64) have $\Delta{(G_{\rm BP}-G_{\rm
    rp})}\ge 0.15$ mag. The number distributions of the color
deviation for both the observed and the simulated stars are shown in
the top panels of Figures \ref{F9a} and \ref{F9b}. These histograms
also demonstrate that the observed color spreads are more extended
than that of the model.

\section{Discussion and Summary}

The double clusters $h$ and $\chi$ Persei may be the youngest clusters
exhibiting eMSTO regions, at an age of only $\sim$14 Myr. The eMSTO
regions in $h$ and $\chi$ Persei are similar to those in other young
clusters in the Magellanic Clouds, which cannot be explained by
photometric dispersions, differential reddening, internal distance
spreads, nor binary contamination. Although both $h$ and $\chi$ Persei
show apparent eMSTO regions, the complex morphology of their upper MS
cannot be explained by multiply aged stellar populations. We found
that even if we assume an extended star formation history lasting up
to $\sim$100 Myr, the extended MS and MSTO regions of $h$ and $\chi$
Persei still cannot be well explained.

We have cross matched, respectively, 51 and 55 upper MS stars with
reliable measurements of projected rotational velocities in $h$ and
$\chi$ Persei from the data of \cite{Huang06a} and \cite{Huang10a}, as
well as three and five emission-line stars. Most emission-line stars
are located on the red side of the MS, which is in line with
observational results reported in previous studies of young LMC
clusters. However, we do not find any obvious correlation between the
colors and projected rotational velocities of the cluster stars.

We compared our observations with a model representing a coeval
stellar population characterized by a large range of stellar rotation
rates. A simple stellar population with different stellar rotation
rates alone cannot fully explain the observed eMSTO regions. The color
difference caused by different stellar rotation rates among upper MS
stars is negligible compared with the observed width of the MSs and
the MSTOs. A statistical analysis shows that the typical color spreads
of the observed stars are more extended than implied by the model.

It is unlikely that $h$ and $\chi$ Persei are characterized by
significant age spreads among their member stars, since their ages and
any possible ranges in age have been strongly constrained on the basis
of observations in other passbands \citep[e.g.,][]{Curr10a}. { In
  addition, we did not find any strong concentration of dust based on
  the IRAS \unit{100}{\micro\meter} images, indicating that the gas
  content in these two clusters is very limited (assuming that the gas
  and dust are spatially co-located).} It thus appears that some
underlying physical phenomena may have changed the specific stellar
colors in the {\sl Gaia} passbands, because the MSs are extended only
for $G\leq 15.5$ mag, which roughly corresponds to stellar masses in
excess of $\sim$1.5--1.6 $M_{\odot}$. This coincides with the mass
range where magnetic braking begins to slow down stellar rotation
\citep{Geor13a}.

The inconsistencies between the morphologies of the eMSTO regions in
the simulated cluster and the observations may be driven by
limitations inherent to the model. As shown by \cite{Nied15a} and
\cite{Bast18a}, the extent of the age spread mimicked by stellar
rotation is a function of the intrinsic cluster age. For 10--20
Myr-old clusters such as $h$ and $\chi$ Persei, the age spread is
expected to be only 2--3 Myr \citep[][their Figure
  4]{Bast18a}. Clearly, the observed width of the eMSTO region would
imply an `age spread' that is much more extended than that
predicted. This may suggest that there could be a problem in the
model's MSTO calibration for stars in very young clusters. For
instance, Be/Oe stars with disks seen edge-on will be reddened by
their own disks, an aspect that is not considered in the model. Stars
that are observed pole-on will show an excess blue color caused by
scattered light off the disk.

\cite{Dant17a} advised that blue MS stars are initially rapidly
rotating stars that have braked recently. These stars are less
advanced in their evolution compared with non-rotating stars with
equivalent masses. They suggest that braking of the blue MS stars may
be caused by tidal torquing of binary components, indicating that blue
upper-MS stars should have binary companions. If this model is valid,
there should be a gradient of stellar rotational velocities across the
full MS, with stars on the blue side being slowly or non-rotating
stars, while stars on the red side would be rapid rotators. Our result
does not exhibit such an apparent correlation between the stellar
colors and their projected rotational velocities. In addition, the
color distribution of binaries and normal stars are similar.

An overlooked role possibly related to the presence of eMSTO regions
is convective overshooting. \cite{Yang17a} calculated the impact of
convective overshooting on the MSTO of clusters as young as 100
Myr. They found that adoption of varying overshooting
parameters($\delta_{\rm ov}$) for individual stars---with $\delta_{\rm
  ov}$ varying from 0.0 to 0.7---can potentially explain the observed
eMSTOs in young and intermediate-age clusters. Indeed, variations in
stellar overshooting parameters might better mimic an age spread than
the presence of different stellar rotation rates, at least for young
clusters. This is because convective overshooting is more efficient
than rotation at bringing H-rich material into the H-burning core,
thus increasing the stellar MS lifetime. In the mean time,
terminal-age MS stars with different $\delta_{\rm ov}$ may show a
spread in nitrogen abundance. This is consistent with some
observations, since the MSTO regions of some young clusters are much
broader when ultraviolet (UV) observations are involved (e.g.,
F336W/F343N passbands of the {\sl Hubble Space Telescope}'s UVIS/WFC3
camera), which are sensitive to nitrogen abundances
\citep[e.g.,][]{Lars14a}. Such a nitrogen abundance dispersion cannot
be produced by any combination of rotating models with moderate
overshooting, $\delta_{\rm ov}=0.2$, and non-rotating models. This may
then be used to differentiate convective core overshooting models from
rotation models \citep{Yang17a}.

Binary interactions may have contributed to form the eMSTO region as
well \citep[e.g.,][]{Yang11a,Yang18a}. These authors claim that some
of the blue MS and MSTO stars in the CMDs of Magellanic Cloud clusters
may be binaries or binary products (including blue straggler stars;
BSSs). Using the cluster NGC 1866 as a testbed, \cite{Yang18a} showed
that a fraction of the blue MS stars can be explained by either merged
stars, MS--naked-He star systems, or MS--white dwarf (WD) systems. In
the case of $h$ and $\chi$ Persei, it is unlikely that there will be a
population of WD stars because of their young ages. MS--WD systems, if
detected, would strongly suggest that these clusters contain a
population of stars that are at least $\sim$30 Myr old, thus
confirming the presence of an internal age spread. The presence of
MS--He star systems or binary merger BSSs is more likely. Stars with
peculiar colors compared with the ZAMS may be stars that are out of
thermal equilibrium and which are probably kept there by being current
mass-transfer systems. These stars can be identified by examining if
they exhibit W Ursae Majoris-like light curves.

% If most stars in the upper MS are merged binaries, 
% then the mass for turnoff stars described by the $\sim$14 Myr isochrone would not exceed 
% twice the mass for turnoff stars of the bulk stellar population. This would indicate a genuine 
% age of up to $\sim$50 Myr for the clusters.

\begin{figure*}
\includegraphics[width=2\columnwidth]{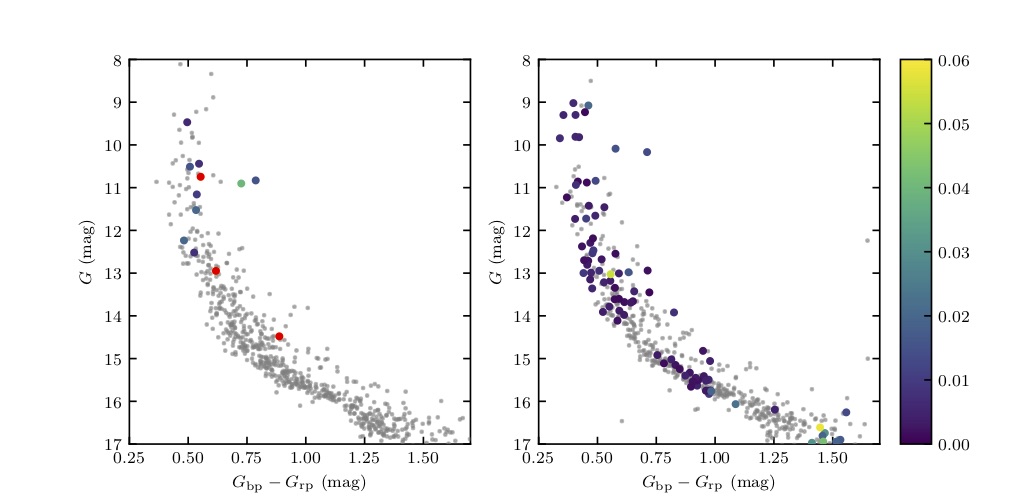}
\caption{CMDs of $h$ (left) and $\chi$ Persei (right) with variable
  stars color-coded by their variability amplitudes. The red dots in
  $h$ Persei represent EA-type eclipsing binaries, which have
  amplitudes of 0.15--0.25 mag.}
\label{F10}
\end{figure*}

Note that variable stars among the clusters' MS stars may affect the
morphology of the CMD. \cite{Sali18a} tested this scenario for the
eMSTO of the cluster NGC 1846 (aged 1--2 Gyr). However, this
hypothesis has not been examined for young clusters. A sample of
pulsating B-type stars has been detected in both $h$ and $\chi$ Persei
\citep[e.g.,][]{Krze99a,Maje08a,Saes10a}. Combined with our ongoing
survey of variable stars in $h$ Persei { using the 50 BiN telescope
  \citep{Xin16a}}, we present the distribution of these variable stars
in the color--magnitude diagrams (see { the left-hand panel} of
Figure \ref{F10}). The colors represent the variability amplitudes,
with the exception of the three red dots in $h$ Persei. The latter are
classified as EA-type eclipsing binaries, and all have variability
amplitudes greater than 0.15 mag. { For $\chi$ Persei, we directly
  use the catalog of variable stars provided by \cite{Saes10a}}. {
  As shown in Figure \ref{F10}}, although the majority of pulsating
stars in these two clusters have relatively small amplitudes, which
are unlikely to explain the widths of their MSs, these eclipsing
binaries could change their location along the CMD fairly easily and
they could therefore be a possible source of broadened MSs. { Note
  that \cite{Saes10a} observed variable stars based on multi-site
  observations with a total exposure time of over 1200 h. The number
  of variable stars detected in $\chi$ Persei is higher than the
  equivalent number in $h$ Presei. Given that $h$ and $\chi$
  Persei are double clusters with almost identical physical
  parameters, we conclude that the genuine number of variable stars in
  $h$ Persei should be much higher than that shown in Figure
  \ref{F10}.}

In summary, the complexity of the MSTO regions in $h$ and $\chi$
Persei may be caused by the combination different underlying physical
processes. { We also suggest that more detailed models of the
  stellar rotation of extremely young stellar populations
  ($\sim$10--20 Myr) are required to fully understand the complex MSTO
  regions in these young clusters.}

\acknowledgements

We thank the anonymous referee for valuable comments. We thank
Dr. Yang Wuming at Beijing Normal University for useful
discussions. We thank Dr. Yu Heng at Beijing Normal University 
for his careful review on details of this work. C. L. was supported by 
the Macquarie Research Fellowship
Scheme. This work was supported by the National Key Research and
Development Program of China through grant 2017YFA0402702 (RdG). This
work was also partly supported by the National Natural Science
Foundation of China through grants 11633005 (L. D.), 11373010, 11473037 and
11803048. K. W. acknowledge funding by the China Scholarship Council. 
This paper has received funding from the European Research 
Council (ERC) under the European Union's Horizon 2020 research 
innovation programme (Grant Agreement ERC-StG 2016, No 716082 
`GALFOR', PI Milone) and from MIUR through the the FARE project 
R164RM93XW ‘SEMPLICE’ (PI Milone).

{\bf This work has made use of data from the European Space Agency (ESA) mission
{\it Gaia} (\url{https://www.cosmos.esa.int/gaia}), processed by the {\it Gaia}
Data Processing and Analysis Consortium (DPAC,
\url{https://www.cosmos.esa.int/web/gaia/dpac/consortium}). Funding for the DPAC
has been provided by national institutions, in particular the institutions
participating in the {\it Gaia} Multilateral Agreement.}

\vspace{5mm} 
\facilities{ESA Gaia, Infrared Astronomical Satellite (IRAS,
  NASA/IPAC)}
  
\software{MIST\citep{Paxt11a,Paxt13a,Paxt15a,Choi16a,Dott16a}, TLUSTY and
SYNSPEC \citep{Hube95a}}

\end{document}